# Exact solution of a kinetic equation of one-step processes of growth and decay

© Ikhtier Holmamatovich Umirzakov

*The laboratory of modeling. Kutateladze Institute of Thermophysics of SB of RAS.*
*Prospect Lavrenteva, 1.Novosibirsk 630090 Russia.*
*Tel.: +7 (383) 354-20-17. E-mail: tepliza@academ.org*

**Abstract**

The exact solution of the kinetic equation of the one step process of a growth (birth) in the presence of sources (runs-off) is found for arbitrary dependence of the rate of the growth on the number of the state (or size). It is shown that the presence of the source leads on the broadening of the state (or size) distribution. It is shown also that the solution obtained can be used for approximate definition of the size distribution of nuclei for the case of large supersaturation of the vapor. The exact solution of the kinetic equation of the one step process of a decay in the presence of sources (runs-off) is found also for arbitrary dependence of the rate of the decay on the number of the state (or size).

# Точное решение кинетического уравнения для одноступенчатого процесса роста и распада

© **Умирзаков Ихтиёр Холмаматович**

*Лаборатория моделирования, Институт теплофизики СО РАН, проспект Лаврентьева, 1, 630090, г. Новосибирск, Россия*
*Тел.: (383) 354-20-17, E-mail: tepliza@academ.org*

**Ключевые слова**: *кинетическое уравнение, процессы роста и распада, источник, сток.*

### Аннотация

Найдено точное решение кинетического уравнения роста (рождения) в присутствии источников (стоков) для произвольной зависимости скорости роста от номера состояния (или размера). Показано, что присутствие источника приводит к уширению распределения по состояниям (или размерам). Показано также, что полученное решение может быть использовано для приближенного определения функции распределения зародышей в случае большого пересыщения пара. Найдено также точное решение кинетического уравнения распада в присутствии источников (стоков) для произвольной зависимости скорости распада от номера состояния (или размера).

## Введение

Рассматривается кинетическое уравнение для одноступенчатого (одношагового) процесса

$$\frac{\partial P_n(t)}{\partial t} = \alpha_{n-1} \cdot P_{n-1}(t) + \beta_{n+1} \cdot P_{n+1}(t) - (\alpha_n + \beta_n) \cdot P_n(t) + g_n(t), \quad (1)$$

где $P_n(t)$ - число объектов в $n$-ом состоянии в момент времени $t$, $\alpha_n$ - частота переходов из $n$-го состояния в $n+1$-ое, $\beta_n$ - частота переходов из $n$-го состояния в $n-1$-ое, а $g_n(t)$ - интенсивность источника (стока) для $n$-го состояния, которая считается заданной функцией времени. Частоты $\alpha_n$ и $\beta_n$ предполагаются независящими от времени; $\alpha_n$ соответствует росту (рождению), а $\beta_n$ - распаду (гибели). Данное уравнение имеет достаточно широкое применение в физике, химии и биологии [1, 2].

Ранее уравнение (1) исследовалось для случая линейной зависимости частот $\alpha_n$ и $\beta_n$ от $n$ (в частности, при $\alpha_n = \beta_n = const$) и отсутствия источников ($g_n(t) = 0$) [2]. В данной работе получено точное решение этого кинетического уравнения для произвольной зависимости $\alpha_n$ (или $\beta_n$) от $n$ в случае процессов, направленных в одну сторону: роста ($\alpha_n \neq 0, \beta_n = 0$) или распада ($\beta_n \neq 0, \alpha_n = 0$).

## Основная часть

1. Рассмотрим случай, когда переходы из $n$-го состояния в $n-1$-ое отсутствуют ($\beta_n = 0$). Тогда (1) принимает вид

$$\frac{\partial P_n(t)}{\partial t} = \alpha_{n-1} \cdot P_{n-1}(t) - \alpha_n \cdot P_n(t) + g_n(t). \quad (2)$$

Это уравнение используется для описания процессов конденсации в случае больших пересыщений пара [3], необратимой коагуляции элементарных точечных дефектов в твердых телах [4], необратимых последовательных мономолекулярных реакций в открытых системах [5] и др.

Предположим, что $g_n(t) = 0$ и $P_n(0) = 0$ для $n < m$, тогда индекс $n$ принимает значения $m, m+1, m+2, \ldots$ . Решение уравнения (2) при произвольно заданном начальном распределении $P_n(0) = P_n(t=0)$ будем искать в виде

$$P_m(t) = f_m(t) \cdot e^{-\alpha_m \cdot t}, \tag{3a}$$

$$P_{m+1}(t) = f_{m+1}(t) \cdot \alpha_m \cdot e^{-\alpha_{m+1} \cdot t}, \tag{3b}$$

$$P_n(t) = f_n(t) \cdot \alpha_m \cdot \ldots \cdot \alpha_{n-1} \cdot e^{-\alpha_n \cdot t}, \quad n = m+2, m+3, \ldots \tag{3c}$$

Подставляя (3) в (2), приходим к следующему уравнению для неизвестной функции $f_n(t)$

$$\frac{\partial f_m(t)}{\partial t} = g_m(t) \cdot e^{\alpha_m \cdot t}, \tag{4a}$$

$$\frac{\partial f_{m+1}(t)}{\partial t} = e^{(\alpha_{m+1} - \alpha_m) \cdot t} \cdot f_m(t) + g_{m+1}(t) \cdot e^{\alpha_{m+1} \cdot t} / \alpha_m, \tag{4b}$$

$$\frac{\partial f_n(t)}{\partial t} = e^{(\alpha_n - \alpha_{n-1}) \cdot t} \cdot f_{n-1}(t) + g_n(t) \cdot e^{\alpha_n \cdot t} / (\alpha_m \cdot \ldots \cdot \alpha_{n-1}), \tag{4c}$$

с начальными условиями

$f_m(0) = P_m(0)$, $f_{m+1}(0) = P_{m+1}(0) / \alpha_m$ и $f_n(0) = P_n(0)/(\alpha_m \cdot \ldots \cdot \alpha_{n-1})$, $n = m+2, m+3, \ldots$.

Интегрируя левую и правую части уравнений (4a)-(4c) по $t$, получаем

$$f_m(t) = f_m(0) + \int_0^t g_m(t) \cdot e^{\alpha_m \cdot t} dt, \tag{4a'}$$

$$f_{m+1}(t) = f_{m+1}(0) + \int_0^t f_m(t) \cdot e^{(\alpha_{m+1} - \alpha_m) \cdot t} dt + \int_0^t g_{m+1}(t) \cdot e^{\alpha_{m+1} \cdot t} dt / \alpha_m, \tag{4b'}$$

$$f_n(t) = f_n(0) + \int_0^t f_{n-1}(t) \cdot e^{(\alpha_n - \alpha_{n-1}) \cdot t} dt + \int_0^t g_n(t) \cdot e^{\alpha_n \cdot t} dt /(\alpha_m \cdot \ldots \cdot \alpha_{n-1}), n = m+2, m+3, \ldots \tag{4c'}$$

Применим к первым интегралам в левых частях уравнений (4b') и (4c') операцию интегрирования по частям:

$$\int_0^t f_m(t) \cdot e^{(\alpha_{m+1} - \alpha_m) \cdot t} dt = \left. \frac{f_m(t) \cdot e^{(\alpha_{m+1} - \alpha_m) \cdot t}}{\alpha_{m+1} - \alpha_m} \right|_0^t - \int_0^t \frac{e^{(\alpha_{m+1} - \alpha_m) \cdot t}}{\alpha_{m+1} - \alpha_m} \cdot \frac{\partial f_m(t)}{\partial t} dt,$$

$$\int_0^t f_{n-1}(t) \cdot e^{(\alpha_n - \alpha_{n-1}) \cdot t} dt = \left. \frac{f_{n-1}(t) \cdot e^{(\alpha_n - \alpha_{n-1}) \cdot t}}{\alpha_n - \alpha_{n-1}} \right|_0^t - \int_0^t \frac{e^{(\alpha_n - \alpha_{n-1}) \cdot t}}{\alpha_n - \alpha_{n-1}} \cdot \frac{\partial f_{n-1}(t)}{\partial t} dt,$$

и подставим в полученные интегралы $\frac{\partial f_m(t)}{\partial t}$ и $\frac{\partial f_{n-1}(t)}{\partial t}$ из (4a') и (4c'). Повторяя эту процедуру $n$-$m$ раз, получаем рекуррентные формулы:

$$f_m(t) = f_m(0) + \int_0^t g_m(t) \cdot e^{\alpha_m \cdot t} dt,$$

$$f_{m+1}(t) = [f_m(t) \cdot e^{(\alpha_{m+1} - \alpha_m) \cdot t} - f_m(0) - \int_0^t e^{\alpha_{m+1} \cdot t} \cdot g_m(t) dt]/(\alpha_{m+1} - \alpha_m) +$$

$$+ \int_0^t e^{\alpha_{m+1} \cdot t} \cdot g_{m+1}(t) dt / \alpha_m + f_{m+1}(0),$$

$$f_{m+2}(t) = [f_{m+1}(t) \cdot e^{(\alpha_{m+2} - \alpha_{m+1}) \cdot t} - f_{m+1}(0) - \int_0^t \frac{e^{\alpha_{m+2} \cdot t} \cdot g_{m+1}(t)}{\alpha_m} dt]/(\alpha_{m+2} - \alpha_{m+1}) -$$

$$- [f_m(t) \cdot e^{(\alpha_{m+2} - \alpha_m) \cdot t} - f_m(0) - \int_0^t e^{\alpha_{m+2} \cdot t} \cdot g_m(t) dt]/[(\alpha_{m+2} - \alpha_{m+1})(\alpha_{m+2} - \alpha_m)] +$$

$$+ \int_0^t \frac{e^{\alpha_{m+2} \cdot t} \cdot g_{m+2}(t)}{\alpha_m \cdot \ldots \cdot \alpha_{m+1}} dt + f_{m+2}(0),$$

$$f_n(t) = \sum_{l=1}^{n-m-2} \frac{(-1)^{l+1}}{\prod_{k=1}^{l}(\alpha_n - \alpha_{n-k})} \cdot [f_{n-l}(t) \cdot e^{(\alpha_n - \alpha_{n-l}) \cdot t} - f_{n-l}(0) - \int_0^t \frac{e^{\alpha_n \cdot t} \cdot g_{n-l}(t)}{\alpha_m \cdot \ldots \cdot \alpha_{n-l-1}} dt] +$$

$$+ \frac{(-1)^{n-m}}{\prod_{k=1}^{n-m-1}(\alpha_n - \alpha_{n-k})} [f_{m+1}(t) e^{(\alpha_n - \alpha_{m+1}) \cdot t} - f_{m+1}(0) - \int_0^t e^{\alpha_n \cdot t} g_{m+1}(x) dx / \alpha_m] +$$

$$+ \frac{(-1)^{n-m+1}}{\prod_{k=1}^{n-m}(\alpha_n - \alpha_{n-k})} [f_m(t) e^{(\alpha_n - \alpha_m) \cdot t} - f_m(0) - \int_0^t e^{\alpha_n \cdot t} g_m(x) dx +$$

$$+ \int_0^t \frac{e^{\alpha_n \cdot t} \cdot g_n(t)}{\alpha_m \cdot \ldots \cdot \alpha_{n-1}} dt + f_n(0), \qquad n = m+3, \; m+4, \; \ldots \; .$$

Если $g_n(t) = I_n e^{-\lambda_n \cdot t}, n = m, \; m+1, \; m+2,\ldots$, где $I_n$ и $\lambda_n$ не зависят от времени, то интегралы в последних четырех равенствах легко вычисляются аналитически, в том числе при наиболее интересных случаях:

- $\lambda_n = 0, \; n = m, \; m+1, \; m+2,\ldots$;
- $\lambda_m \neq 0, \; I_m \neq 0, \; I_n = 0, \; n = m+1, \; m+2,\ldots$;
- $\lambda_m = 0, \; I_m \neq 0, \; I_n = 0, \; n = m+1, \; m+2,\ldots$ .

2. Рассмотрим времена $t << \min[1/|\alpha_{n-1} - \alpha_n|, \; n = m+1, \; m+2,\ldots]$. Тогда экспоненты при $f_m(t)$ и $f_{n-1}(t)$ в (4а)-(4с) можно считать равными единице. В этом можно убедиться прямой подстановкой полученного решения в (4а)-(4с). В результате (4) принимает вид

$$\frac{\partial f_m(t)}{\partial t} = g_m(t) \cdot e^{\alpha_m \cdot t},$$

$$\frac{\partial f_{m+1}(t)}{\partial t} = f_m(t) + g_{m+1}(t) \cdot e^{\alpha_{m+1} \cdot t} / \alpha_m,$$

$$\frac{\partial f_n(t)}{\partial t} = f_{n-1}(t) + g_n(t) \cdot e^{\alpha_n \cdot t} /(\alpha_m \cdot \ldots \cdot \alpha_{n-1}).$$

Действуя по приведенной выше схеме получаем

$$f_m(t) = f_m(0) + \int_0^t g_m(t) \cdot e^{\alpha_m \cdot t} dt, \tag{5a}$$

$$f_{m+1}(t) = t \cdot f_m(t) - \int_0^t t \cdot e^{\alpha_m \cdot t} \cdot g_m(t) dt + \int_0^t e^{\alpha_{m+1} \cdot t} \cdot g_{m+1}(t) dt / \alpha_m + f_{m+1}(0), \tag{5b}$$

$$f_{m+2}(t) = t \cdot f_{m+1}(t) - \int_0^t t \cdot e^{\alpha_{m+1} \cdot t} \cdot g_{m+1}(t)dt/\alpha_m - t^2 \cdot f_m(t)/2 + \int_0^t t^2 \cdot e^{\alpha_m \cdot t} \cdot g_m(t)dt +$$

$$+ \int_0^t e^{\alpha_{m+2} \cdot t} \cdot g_{m+2}(t)dt/\alpha_m \alpha_{m+1} + f_{m+2}(0), \quad (5c)$$

$$f_n(t) = \sum_{l=1}^{n-m-2} \frac{(-1)^{l+1}}{l!} \cdot [t^l \cdot f_{n-l}(t) - \int_0^t t^l \cdot \frac{e^{\alpha_{n-l} \cdot t} \cdot g_{n-l}(t)}{\alpha_m \cdot \ldots \cdot \alpha_{n-l-1}} dt] +$$

$$+ \frac{(-1)^{n-m}}{(n-m-1)!}[t^{n-m-1} \cdot f_{m+1}(t) - \int_0^t t^{n-m-1} \cdot \frac{e^{\alpha_{m+1} \cdot t} \cdot g_{m+1}(t)}{\alpha_m} dt] + \quad (5d)$$

$$+ \frac{(-1)^{n-m+1}}{(n-m)!}[t^{n-m} \cdot f_m(t) - \int_0^t t^{n-m} \cdot e^{\alpha_m \cdot t} \cdot g_m(t)dt] + \int_0^t \frac{e^{\alpha_n \cdot t} \cdot g_n(t)}{\alpha_m \cdot \ldots \cdot \alpha_{n-1}} dt + f_n(0),$$

$n = m+3, \ m+4, \ldots \ .$

    2.1. В случае, когда $g_n(t) = I_n e^{-\lambda_n \cdot t}$, $P_n(0) = N_n$, $\alpha_n \neq \lambda_n$, $n = m, \ m+1, \ m+2, \ldots$, где $I_n$, $\lambda_n$ и $N_n$ не зависят от времени, в силу линейности уравнений (2) имеем

$$P_n(t) = \sum_{L=m}^{n} \alpha_L \cdot \ldots \cdot \alpha_{n-1} \cdot e^{-\alpha_n \cdot t} \cdot \frac{N_L \cdot t^{n-L}}{(n-L)!} + \sum_{i=m}^{n} \frac{I_i}{[(\alpha_i - \lambda_i) \cdot t]^{n-i+1}} \cdot [e^{(\alpha_i - \lambda_i) \cdot t} - \sum_{k_i=0}^{n-i} \frac{[(\alpha_i - \lambda_i) \cdot t]^{k_i}}{k_i!}]],$$

$n = m, \ m+1, \ m+2, \ldots \ .$

    2.2. Пусть $g_m(t) = I = const$, $P_m(0) = N = ñonst$, $g_n(t) = 0$, $P_n(0) = 0$, $n = m+1, \ m+2, \ \ldots$. Тогда (5a)-(5d) записываются в виде

$$f_n(t) = N \cdot \frac{t^{n-m}}{(n-m)!} + \frac{I}{\alpha_m^{n-m+1}} \cdot [e^{\alpha_m \cdot t} - \sum_{k=0}^{n-m} \frac{(\alpha_m \cdot t)^k}{k!}], \quad n = m, \ m+1, \ m+2, \ldots \ . \quad (6)$$

Тогда с учетом (3) из (6) окончательно имеем

$$P_n(t) = \alpha_m \cdot \ldots \cdot \alpha_{n-1} \cdot e^{-\alpha_n \cdot t} \cdot [\frac{N \cdot t^{n-m}}{(n-m)!} + \frac{I}{\alpha_m^{n-m+1}} \cdot [e^{\alpha_m \cdot t} - \sum_{k=0}^{n-m} \frac{(\alpha_m \cdot t)^k}{k!}]], \quad n = m, \ m+1, \ m+2, \ldots \ . \quad (7)$$

При использовании формулы (7) удобно применить равенство

$$e^x - \sum_{k=0}^{n} \frac{x^k}{k!} = \frac{1}{n!} \cdot \int_0^x e^y \cdot (x-y)^n dy. \quad (8)$$

Оно легко доказывается с помощью принципа математической индукции [6, стр.558]. Действительно, при $n = 0$ и $n = 1$ оно удовлетворяется. Предположим, что оно верно для произвольного $n$ ($n > 1$). Рассмотрим левую часть равенства (8) для $n+1$. Используя (8) для $n$ и интегрирование по частям получаем

$$e^x - \sum_{k=0}^{n+1} \frac{x^k}{k!} = e^x - \sum_{k=0}^{n} \frac{x^k}{k!} - \frac{x^{n+1}}{(n+1)!} = \frac{1}{n!} \cdot \int_0^x e^y \cdot (x-y)^n dy - \frac{x^{n+1}}{(n+1)!} = -\frac{e^y}{n!} \cdot \frac{(x-y)^{n+1}}{\cdot(n+1)}\Big|_0^x +$$

$$+ \int_0^x \frac{e^y}{n!} \cdot \frac{(x-y)^{n+1}}{\cdot(n+1)} dy - \frac{x^{n+1}}{(n+1)!} = \frac{1}{(n+1)!} \cdot \int_0^x e^y \cdot (x-y)^{n+1} dy.$$

    В качестве одного из приложений формулы (7) можно указать на конденсацию сильно пересыщенного пара в изотермической системе (в этом случае можно пренебречь испарением кластеров) [3]. Ввиду предполагаемого в (7) отсутствия зависимости $\alpha_n$ от $t$ в качестве дополнительного предположения здесь следует еще принять, что доля конденсата мала, так что истощением мономеров можно пренебречь. В указанных условиях (7) будет описывать эволюцию функции распределения кластеров $P_n(t)$ по размерам $n$ после того, как в момент $t=0$ в системе создали большое пересыщение. При

этом $I$ равна скорости зародышеобразования и нужно положить $n = g - g_c$, где $g$ - число частиц (атомов или молекул) в зародыше новой фазы, $g_c$ - размер критического зародыша.

Для определения функции распределения кластеров $P_n(t)$ по размерам в явном виде требуется точное знание зависимости свободной энергии (фазового объема) кластера от температуры (энергии) и размера кластера. Определение этой зависимости представляет собой чрезвычайно сложную задачу с теоретической, вычислительной и экспериментальной точек зрения (смотрите, например, [7-16]). Знание распределения кластеров по размерам необходимо в разработке нанотехнологий и для практического применения [14-16].

Рассмотрим случай $\alpha_n = \alpha = const$, $I > 0$ (источник). При этом ограничение $t \ll \min[1/|\alpha_{n-1} - \alpha_n|, n = m+1, m+2, ....]$ на $t$ снимается и (7) является точным решением (3). Если для простоты обозначений положить $m=0$, то с учетом (8) оно принимает вид

$$P_n(t) = N \cdot \frac{(\alpha \cdot t)^n}{n!} \cdot e^{-\alpha \cdot t} + \frac{I}{\alpha \cdot n!} \cdot e^{-\alpha \cdot t} \cdot \int_0^{\alpha \cdot t} e^y \cdot (\alpha \cdot t - y)^n dy, \quad n = 0, 1, 2, ... \qquad (9)$$

и при $I=0$ переходит в пуассоновское распределение. В том, что (9) является точным решением кинетического уравнения роста (3), можно убедиться непосредственной подстановкой (9) в (3).

Вычислим среднее значение $<n>$ и квадрат дисперсии $\sigma^2 = <n^2> - <n>^2$, где

$$<n> = \sum_{n=0}^{\infty} n \cdot P_n / \sum_{n=0}^{\infty} P_n,$$

$$<n^2> = \sum_{n=0}^{\infty} n^2 \cdot P_n / \sum_{n=0}^{\infty} P_n.$$

Далее найдем

$$\sum_{n=0}^{\infty} P_n = N e^{-\alpha \cdot t} \sum_{n=0}^{\infty} (\alpha \cdot t)^n / n! + I e^{-\alpha \cdot t} \int_0^{\alpha \cdot t} e^y \sum_{n=0}^{\infty} (\alpha \cdot t - y)^n dy / n! \alpha =$$

$$= N + I e^{-\alpha \cdot t} \int_0^{\alpha \cdot t} e^y e^{\alpha \cdot t - y} dy / \alpha = N + I \cdot t.$$

Аналогично

$$\sum_{n=0}^{\infty} n \cdot P_n = N \cdot \alpha \cdot t + I \cdot \alpha \cdot t^2 / 2,$$

$$\sum_{n=0}^{\infty} n^2 \cdot P_n = N \cdot \alpha \cdot t \cdot (1 + \alpha \cdot t) + I \cdot \alpha \cdot t^2 (\alpha \cdot t / 3 + 1/2).$$

И окончательно:

$$<n> = \alpha \cdot t \cdot (N + I \cdot t / 2) / (N + I \cdot t),$$

$$\sigma^2 = \alpha \cdot t \cdot \frac{N \cdot (1 + \alpha \cdot t) + I \cdot t \cdot (\alpha \cdot t / 3 + 1/2)}{N + I \cdot t} - \alpha^2 \cdot t^2 \cdot \frac{(N + I \cdot t / 2)^2}{(N + I \cdot t)^2},$$

$$\frac{\sigma^2}{<n>^2} = \frac{[N \cdot (1 + \alpha \cdot t) + I \cdot t \cdot (\alpha \cdot t / 3 + 1/2)] \cdot (N + I \cdot t)}{\alpha \cdot t \cdot (N + I \cdot t / 2)^2} - 1.$$

Легко показать, что среднее значение и дисперсия увеличиваются с увеличением скорости роста $\alpha$, а квадрат отношения дисперсии к среднему значению уменьшается с увеличением $\alpha$.

В предельном случае $t \ll N/I$ получаем соотношения:

$$<n> = \alpha \cdot t, \quad \sigma^2 = \alpha \cdot t \cdot (1 + \alpha \cdot t) - \alpha^2 \cdot t^2 = \alpha \cdot t, \quad \frac{\sigma^2}{<n>^2} \approx \frac{1}{\alpha \cdot t} \approx \frac{1}{<n>}.$$

Эти соотношения становятся точными при *I*=0. Первые два соотношения дают известный результат для пуассоновского распределения: $\sigma^2 = <n>$.

При $t \gg N/I$ получаем соотношения, становящиеся точными при *N*=0:
$<n> = \alpha \cdot t / 2, \quad \sigma^2 = \alpha \cdot t \cdot (1/2 + \alpha \cdot t / 3) - \alpha^2 \cdot t^2 / 4 = \alpha \cdot t / 2 + \alpha^2 \cdot t^2 / 12,$

$$\frac{\sigma^2}{<n>^2} \approx \frac{1}{3} + \frac{2}{\alpha \cdot t} \approx \frac{1}{3} + \frac{1}{<n>}.$$

Отсюда видно, что наличие интенсивного источника приводит к уширению функции распределения, причем таким образом, что при $<n> \to \infty$ отношение $\sigma / <n>$ сохраняет конечную величину, равную $1/\sqrt{3}$.

**3.** Аналогично может быть исследован случай, когда переходы из *n*-го состояния в *n*+1−ое состояние отсутствуют (нет роста или рождения: $\alpha_n = 0$). При этом (1) принимает вид

$$\frac{\partial P_n(t)}{\partial t} = \beta_{n+1} \cdot P_{n+1}(t) - \beta_n \cdot P_n(t) + g_n(t). \tag{10}$$

В этом уравнении мы предполагаем, $g_n(t) = 0$ и $P_n(0) = 0$ для $n > m$, и поэтому $n = m, m-1, m-2, \ldots$ . Решение уравнения (10) при произвольно заданном начальном распределении ищем в виде

$P_m(t) = f_m(t) \cdot e^{-\beta_m \cdot t}, \qquad P_{m-1}(t) = f_{m-1}(t) \cdot \beta_m \cdot e^{-\beta_{m-1} \cdot t},$
$P_n(t) = f_n(t) \cdot \beta_{n+1} \cdot \ldots \cdot \beta_m \cdot e^{-\beta_n \cdot t}, \qquad n = m-2, m-3, \ldots$ .

Действуя аналогично п. 1, для $f_n(t)$ получаем следующие рекуррентные формулы

$$f_m(t) = f_m(0) + \int_0^t g_m(t) \cdot e^{\beta_m \cdot t} dt,$$

$$f_{m-1}(t) = [f_m(t) \cdot e^{(\beta_{m-1} - \beta_m) \cdot t} - f_m(0) - \int_0^t e^{\beta_{m-1} \cdot t} \cdot g_m(t) dt]/(\beta_{m-1} - \beta_m) +$$

$$+ \int_0^t e^{\beta_{m-1} \cdot t} \cdot g_{m-1}(t) dt / \beta_m + f_{m-1}(0),$$

$$f_{m-2}(t) = [f_{m-1}(t) \cdot e^{(\beta_{m-2} - \beta_{m-1}) \cdot t} - f_{m-1}(0) - \int_0^t e^{\beta_{m-2} \cdot t} \cdot g_{m-1}(t) dt / \beta_m]/(\beta_{m-2} - \beta_{m-1}) -$$

$$- [f_m(t) \cdot e^{(\beta_{m-2} - \beta_m) \cdot t} - f_m(0) - \int_0^t e^{\beta_{m-2} \cdot t} \cdot g_m(t) dt]/[(\beta_{m-2} - \beta_{m-1})(\beta_{m-2} - \beta_m)] +$$

$$+ \int_0^t e^{\beta_{m-2} \cdot t} \cdot g_{m-2}(t) dt / \beta_m \beta_{m-1} + f_{m-2}(0),$$

$$f_n(t) = \sum_{l=1}^{m-n-2} \frac{(-1)^{l+1}}{\prod_{k=1}^{l}(\beta_n - \beta_{n+k})} \cdot [f_{n+l}(t) \cdot e^{(\beta_n - \beta_{n+l}) \cdot t} - f_{n+l}(0) - \int_0^t \frac{e^{\beta_n \cdot t} \cdot g_{n+l}(t)}{\beta_{n+l+1} \cdot \ldots \cdot \beta_m} dt] +$$

$$+ \frac{(-1)^{m-n}}{\prod_{k=1}^{m-n-1}(\beta_n - \beta_{n+k})} [f_{m-1}(t) \cdot e^{(\beta_n - \beta_{m-1}) \cdot t} - f_{m-1}(0) - \int_0^t \frac{e^{\beta_n \cdot t} \cdot g_{m-1}(t)}{\beta_m} dt] +$$

$$+ \frac{(-1)^{m-n+1}}{\prod_{k=1}^{m-n}(\beta_n - \beta_{n+k})} [f_m(t) e^{(\beta_n - \beta_m) \cdot t} - f_m(0) - \int_0^y e^{\beta_n \cdot t} g_m(t) dt] +$$

$$+ \int_0^t \frac{e^{\beta_n \cdot t} \cdot g_n(t)}{\beta_{n+1} \cdot \ldots \cdot \beta_m} dt + f_n(0), \qquad n = m-3, \ m-4, \ \ldots .$$

Предел $t << \min[1/|\beta_{n+1} - \beta_n|, \ n = m-1, \ m-2, \ldots]$ рассматривается аналогичным пункту 2 образом.

Если решать кинетические уравнения (2) и (10) численными методами, то, как легко увидеть, для определения $P_n(t)$ необходимо численно вычислить многомерные интегралы кратности от 1 до $|n-m|+1$. При этом с увеличением кратности интеграла быстро увеличиваются ошибки расчетов, что приводит к трудно контролируемым ошибкам при $|n-m|+1 >> 1$. Поэтому необходимым становиться использование полученных в настоящей работе результатов.

**Выводы**

1. Точное решение кинетического уравнения роста (рождения) в присутствии произвольно зависящих от времени источников (стоков) для произвольной зависимости скорости роста (постоянной во времени) и источников (стоков) от номера состояния (или размера) для произвольных начальных условий может быть найдено в замкнутом виде. В случае постоянных во времени источников (стоков) решение имеет аналитический вид.
2. Присутствие источника приводит к уширению распределения по состояниям (или размерам).
3. Точное решение кинетического уравнения распада в присутствии произвольно зависящих от времени источников (стоков) для произвольной зависимости скорости распада (постоянной во времени) и источников (стоков) от номера состояния (или размера) для произвольных начальных условий может быть найдено в замкнутом виде. В случае постоянных во времени источников (стоков) решение имеет аналитический вид.
4. Полученные в работе решения необходимы как при аналитическом, так и при численном исследовании функции распределения как для процессов роста (рождения) и так и процесса распада в присутствии источников (стоков).

**Литература**